**Excessive Screen Time is Associated with Mental Health Problems and ADHD in US Children and Adolescents: Physical Activity and Sleep as Parallel Mediators**

Ying Dai, PhD [1], Na Ouyang, PhD candidate, MSN [2*]

[1] Department of Nursing, Guangzhou Women and Children's Medical Center, Guangzhou Medical University, Guangzhou, China

[2] School of Nursing, Yale University, Orange, CT, USA

*Correspondence to:

Na Ouyang, PhD candidate, MSN
School of Nursing
Yale University
400 West Campus Drive
Orange, CT 06477
Email: na.ouyang@yale.edu

# Excessive Screen Time is Associated with Mental Health Problems and ADHD in US Children and Adolescents: Physical Activity and Sleep as Parallel Mediators

**Abstract**

Objectives: To explore the relationships between screen time and child and adolescent anxiety, depression, behavior/conduct problems, and ADHD during the pandemic, and whether physical activity, sleep duration, and bedtime regularity mediate these relationships.

Methods: We analyzed data from 50231 children and adolescents aged 6-17 years from the US National Survey of Children's Health 2020-2021. We used the exact natural effect and structural equation modeling to investigate the mediating effect of physical activity, short sleep duration, and irregular bedtime on the relationship between screen time and child and adolescent mental health and ADHD.

Results: Daily screen time ≥ 4 hours was associated with higher risks of anxiety (aOR = 1.45, 95% CI: 1.32, 1.58), depression (aOR = 1.65, 95% CI: 1.41, 1.93), behavior/conduct problems (aOR = 1.17, 95% CI: 1.05, 1.30), and ADHD (aOR = 1.21, 95% CI: 1.11, 1.33), respectively. Physical activity mediated 30.2% - 39.3% of the association, followed by irregular bedtime (18.2% - 25.7%), and short sleep duration (2.77% - 7.34%).

Conclusions: Prolonged screen time was associated with poorer mental health and ADHD through the mediation of physical activity, bedtime regularity, and short sleep duration.

Policy implications: Interventions should promote physical activity, regular sleep routines, and adequate sleep duration to effectively mitigate mental health issues and ADHD among children and adolescents.

## Introduction

In 2023, 22% of the US population, approximately 72.8 million, are children aged 0 to 17 years old [1]. Healthy People 2030 sets data-driven national objectives to promote healthy physical, mental, emotional, and behavioral development in children and adolescents [2]. During the COVID-19 pandemic, children experienced social distancing and school lockdowns, which impacted their mental health and health behaviors [3]. The prevalence of mental health disorders has increased during the pandemic. Specifically, between 2016 and 2020, there were significant increases in anxiety (7.1% to 9.2%), depression (3.1% to 4.0%), and behavior or conduct problems (6.7% to 8.1%) in the US [4]. Children and adolescents' screen time increased by 52% during the pandemic [5], while the prevalence of daily physical activity significantly decreased from 24.2% to 19.8 [4], and sleep duration decreased with increased sleep disturbance prevalence [6].

The associations between screen time and mental health or attention-deficit/hyperactivity disorder (ADHD) remain inconsistent. While most studies have found significant positive correlations between screen time and children's internalizing behaviors, such as depression and anxiety [7,8], some studies identified no associations or even negative associations [9,10]. Similarly, there are inconsistent findings on associations between screen time and externalizing behaviors (e.g., ADHD, conduct behaviors). Uddin & Hasan (2023) found that the association between screen use and ADHD was insignificant after controlling sociodemographic and neighborhood factors and resilience. By contrast, a systematic review and meta-analysis found that excessive screen time was significantly associated with externalizing behaviors (e.g., aggression, inattention), even though the correlation coefficient was small [11]. Notably, the COVID-19 pandemic has magnified the role of screen time in children's daily life. On the one hand, various news regarding COVID-19, negative attitudes and emotions towards the pandemic and related restrictions, and social comparison with peers and influencers negatively impacted children's mental health [12]. On the other hand, using social media and the internet was the only way to stay in contact with friends and family members to cope with the absence of direct contact and negative emotions during the pandemic, which could potentially protect their mental health [12]. Owing to the inconsistent findings and behavioral changes during the pandemic, assessing the associations between screen time and mental health or ADHD during the pandemic is needed.

Physical activity and sleep (sleep duration and irregular bedtime) may mediate the associations between screen time and mental health or ADHD, although the roles of physical activity and sleep as mediators are unclear. Excessive screen time could have negative impacts on physical activity and sleep [13]. Engagement in physical activity and having enough sleep have been identified as significant factors related to children's mental health and ADHD [14]. To our knowledge, the mediation effect of sleep on the associations between screen time and mental health remains inconsistent. Specifically, some studies found sleep duration did not mediate the associations between screen time and depression and anxiety [7,8], while other studies found sleep duration mediated the relationship between screen time and problem behaviors, even though the effect sizes were small [15]. However, sleep is a multi-dimensional concept, which includes sleep duration, sleep continuity or efficiency, timing, alertness/sleepiness, and satisfaction/quality [16]. Irregular bedtime has been associated with excessive screen time and low sleep quality [17]. Bedtime variability could predict internalizing and externalizing behaviors [18]. What is notably missing from the literature is the exploration of the mediation role of irregular bedtime and physical activity on associations between screen time and mental health and ADHD.

To address these knowledge gaps, this paper aimed to investigate (1) the relationship between screen time and mental health problems and ADHD in US children and adolescents using the National Survey of Child Health 2020-2021 data; (2) whether physical activity and sleep (short sleep duration and irregular bedtime) mediate these relationships. Based on previous literature, the following hypotheses were derived (see supplementary Figure 1). First, longer screen time is associated with higher risks of mental health problems and ADHD. Second, physical activity, short sleep duration, and irregular bedtime separately and parallelly mediate the relationships between screen time and mental health problems and ADHD.

## Methods

**Participants and study design**

This study used the latest two-year combined National Survey of Child Health (NSCH) dataset, encompassing data from NSCH 2020, and 2021. The NSCH is a nationally representative cross-sectional survey of parents of children aged 0 - 17 years in the US aimed at investigating various aspects of US children and family health outcomes. The survey was initiated in 2003, with subsequent surveys in 2007, 2011/12, and annual surveys from 2016 onwards. The NSCH 2020 was conducted from July 2020 to January 2021, with a total of 42,777 parents/caregivers of children completed the survey, and the NSCH 2021 was conducted from July 2021 to January 2022, with 50,892 parents/caregivers completed the survey. Detailed information about NSCH2020-2021 was reported elsewhere [19]. The data used in this analysis was available at the United States Census Bureau: https://www.census.gov/programs-surveys/nsch/data/datasets.html. A total of 50,231 children aged 6-17 years with complete data on the variables used in this study were included in the analysis.

**Measures**

***Screen time***: Children's screen time was evaluated by asking parents/caregivers about the child's electronic device use on weekdays by stating, "On most weekdays, about how much time does this child usually spend in front of a TV, computer, cellphone or other electronic device watching programs, playing games, accessing the internet or using social media, not including school work?" The question has the following answers: 1 = less than 1 hour per day; 2 = 1 hour per day; 3 = 2 hours; 4 = 3 hours per day; and 5 = 4 or more hours per day. Following previous literature [8], we categorized screen time into light digital media user (< 2 hours per day), moderate user (2-3 hours per day), and heavy user (≥ 4 hours per day).

***Physical activity***: Children's physical activity status was measured by asking parents/caregivers the following question: "During the past week, on how many days did this child exercise, play a sport, or participate in physical activity for at least 60 minutes, age 6 – 17 years?", with the four answers: 0 days; 1 - 3 days; 4 – 6 days; and every day.

***Short sleep duration***: Children's sleep duration was assessed by asking parents/caregivers the following question: "During the past week, how many hours of sleep did this child get on most weeknights?", with seven answers: less than 6 hours; 6 hours; 7 hours; 8 hours; 9 hours; 10 hours; and 11 or more hours. A new variable named short sleep duration was generated based on the appropriate sleep duration recommended by the American Academy of Sleep Medicine [20]. Specifically, children aged 6 - 12 years who slept less than nine hours and children aged 13 - 17 years who slept less than eight hours were defined as having short sleep duration.

***Irregular bedtime:*** Children's bedtime regularity was collected by asking parents/caregivers, "How often does this child go to bed at about the same time on weeknights?" The answers were "1 = Always; 2 = Usually; 3 = Sometimes; 4 = Rarely or Never".

***Mental health problems***: Children's mental and developmental health was measured by asking parents/caregivers, "Has a doctor, other health care provider, or educator ever told you that this child has the following problems including Tourette Syndrome, anxiety problems, depression, behavioral and conduct problem, developmental delay, intellectual disability, speech or other language disorder, learning disability, Autism or Autism Spectrum Disorder, Attention Deficit Disorder or Attention-Deficit/Hyperactivity Disorder (ADHD)?" Parents/caregivers separately answered these questions with "yes/no" answers. If parents/caregivers answered "yes" to this question, they were further asked another question "If yes, does this child currently have the condition?" with "yes/no" answers. Only children whose parents/caregivers answered "yes" to both questions were classified as having the condition. This study only included anxiety problems, depression, behavioral/conduct problems, and ADHD as the main outcomes.

**Covariates**: The following eleven variables were treated as covariates based on previous research findings [8,21]: survey year, child age in years, child gender, ethnicity, insurance type, whether the child was born in the US, primary language spoken at home, parent/caregiver highest education level, household income level, family structure, and family resilience and connection index (FRCI). The FRCI was calculated in accordance with a previous study [8]. The FRCI comprises six questions designed to assess family resilience and connections: when the family is faced with problems, how often are you likely to do each of the following (1) talk together about what to do, (2) work together to solve problems, (3) know we have strengths to draw on, (4) stay hopeful even in difficult times; and how well do parents/caregivers think you can (5) share ideas or talk about things that really matter, and (6) handle the day-to-day demands of raising children. The answers to the first four questions were coded as 0 = none/some of the time, 1 = most of the time, and 2 = all of the time. The answers to the last two questions were coded as 0 = not very well or not very well at all, 1 = somewhat well, and 2 = very well. The total score of the six questions constitutes the FRCI, with higher scores indicating greater resilience and stronger connections with the family.

**Statistical analysis**

Means and standard deviations and frequencies and proportions were used to describe continuous and categorical demographic characteristics, as appropriate. Logistic regression was conducted to obtain odds ratios (OR) for the association between screen time and children's mental health and ADHD, respectively.

The simple mediating effects of physical activity, short sleep duration, and irregular bedtime were separately investigated using the "ExactMed" package [22], which is an extension of Baron and Kenny's method [23]. Screen time was entered as a predictor, and each mental health outcome and ADHD were separately entered as an outcome variable. The natural direct effect (NDE) and natural indirect effects (NIE) of each mediator for children and adolescent mental health and ADHD outcome were calculated using their respective logistic models with 95% bootstrap confidence intervals (CI) in 1000 replications. In simple mediation analyses, the proportion of mediating effects of each mediator was calculated following the inverse odds weighting (IOW) approach with the formula: $\frac{OR^{NDE}*(OR^{NIE}-1)}{OR^{NDE}*OR^{NIE}-1} * 100$ [24].

The parallel mediating effects of physical activity, short sleep duration, and irregular bedtime were tested using the Structural Equation Modelling (SEM) approach with the "LAVAAN" package [25]. Standardized parameters for each path and its 95% confidence intervals with 1000 bootstrap replicates were calculated with the diagonal weighted least squares (DWLS) approach. The proportion of each mediator in parallel mediation analyses was calculated using the following formula: the sum of all indirect effects/total effects*100 [26]. All mediation analyses were adjusted for the above eleven covariates. Screen time, physical activity, short sleep duration, bedtime regularity, child mental health and ADHD, were treated as categorical or binary variables in all mediation models, as appropriate. Given that different age groups have different circadian rhythm development and physical activity levels, multi-group parallel mediation was further conducted to explore how the mediating effects of physical activity and sleep on child mental health and ADHD differ across different age groups (6 - 10 years, 11 - 13 years, and 14 - 17 years). A p-value < 0.05 was considered statistically significant. All analyses were conducted in the statistical software R (version 4.4.0).

## Results

### Demographic characteristics

Supplementary Table 1 shows the detailed descriptive characteristics of children and their household information. A total of 31.7% of children had at least 4 hours of daily screen time, and 21.6% of children had daily physical activities ≥ 60 minutes. Regarding sleep patterns, 34.4% of children had adequate sleep duration, and 24.7% of children always went to bed at the same time during weekdays. The survey-weighted proportion of current anxiety, depression, behavior/conduct problems, and ADHD were 11.1% (95% CI: 10.6 – 11.6%), 4.9% (95% CI: 4.5 – 5.3%), 8.5% (95% CI: 7.9 – 9.0%), and 11.1% (95% CI: 10.5 – 11.6%), respectively.

### Association between screen time and mental health and ADHD

Univariate logistic regression found that screen time was positively associated with anxiety, depression, behavior/conduct problems, and ADHD (supplementary Table 2, unadjusted models). In adjusted models, the effects of screen time were attenuated but remained statistically significant when covariates were included (supplementary Table 2, adjusted models). Heavy screen time (≥ 4 hours per day) was positively associated with worse mental health and ADHD in children and adolescents, with adjusted ORs ranging between 1.17 for behavior/conduct problems and 1.65 for depression.

**Simple mediation modeling**

After controlling for covariates, simple mediation analyses show that physical activity generally exhibits the strongest mediating effects on the relationship between screen time and mental health and ADHD, followed by irregular bedtime, and short sleep duration (supplementary Table 3). For example, 33.2% of the total effect of screen time on adolescent anxiety is mediated through physical activity (OR for natural indirect effect = 1.31, 95% bootstrap CI: 1.25 - 1.36), while the mediating proportions of short sleep duration and irregular bedtime in this association are approximately 2.77% and 18.2%, respectively.

**Parallel mediation modeling**

The mediation effects of physical activity and sleep on the relationship between screen time and anxiety are shown in Figure 1. The indirect effects of the pathways through physical activity ($a_1*b_1 = 0.033$, 95% bootstrap CI: 0.028, 0.039), and irregular bedtime ($a_3*b_3 = 0.031$, 95% bootstrap CI: 0.023, 0.040) are significant mediators in the relation between screen time and anxiety. The pathway through short sleep duration was not significant ($a_2*b_2 = 0.001$, 95% bootstrap CI: -0.001, 0.004). These results indicate that physical activity and irregular bedtime both partially mediate the relationship between screen time and anxiety.

Regarding screen time and depression, behavior/conduct problems, and ADHD, the structural equation modeling consistently showed that the indirect effects in all parallel mediation models were statistically significant (see Figure2, Figure 3, and Figure 4). This indicates that physical activity, short sleep duration, and irregular bedtime were parallel mediators for the associations of screen time with mental health and ADHD. These mediators jointly explained 25.65% to 54.63% of the association between screen time and depression and behavior/conduct problems, respectively.

**Multi-group parallel mediation modeling**

Multi-group mediation analysis reveals that across the three age groups (6 - 10 years, 11 - 13 years, and 14 - 17 years), screen time consistently reduces physical activity and sleep duration, and increases irregular bedtime (Supplementary Figure 2, Supplementary Figure 3, Supplementary Figure 4, and Supplementary Figure 5, path coefficients for a1 and a2). Both physical activity and irregular bedtime consistently mediate the relationship between screen time and mental health and ADHD (Supplementary Figure 2, Supplementary Figure 3, Supplementary Figure 4, and Supplementary Figure 5, path coefficients for a1*b1, and a3*b3). Physical activity generally has a protective effect against poor mental health and ADHD, whereas irregular bedtime exacerbates these conditions. Compared to physical activity and bedtime, the mediating effects of sleep duration on screen time and mental health and ADHD were substantially smaller across all age groups and even became non-significant in the 11 - 13 years age group (Supplementary Figure 2, Supplementary Figure 3, Supplementary Figure 4, and Supplementary Figure 5, path coefficients for a2*b2). The direct effect of screen time on anxiety is significant and positive across all groups, with the total effect most pronounced in the oldest group (14 - 17 years), followed by the middle (11 - 13 years) and youngest (6 - 10 years) groups. However, the direct effects of screen time on depression, behavior/conduct problems, and ADHD were only significant in the oldest age group (14 - 17 years). Supplementary Table 4 showed that compared to multi-group mediation models with no constrain (i.e., full models), assigning equal loading to all age groups (i.e., constrained models) resulted in poorer model fit indices, indicating that multi-group mediation provided a better fit to the data.

**Discussion**

Using the latest population-representative US NSCH 2020-21 dataset, this study aimed to explore the mediating role of physical activity, sleep duration, and bedtime regularity in the relationship between screen time and mental health and ADHD among children and adolescents. The findings showed that screen use was positively associated with worse mental health and ADHD. Physical activity, short sleep duration, and irregular bedtime were parallel mediators in this relationship, in which physical activity had the most profound mediation effects.

**Direct effects of screen time on mental health and ADHD**

Consistent with prior studies using earlier NSCH datasets [7,8], the current study found that screen time use was positively associated with child and adolescent anxiety and depression. Several potential mechanisms may explain this relationship. From a framing theory perspective, children who excessively use digital devices are more likely to be exposed to negatively framed news and information during the pandemic, which could further negatively affect their mental health [12,27]. From the social comparison theory perspective, excessive screen time may lead to the positivity bias in which children and adolescents perceive their peers as happier and less affected by the pandemic, which may exacerbate their own depressive symptoms [12,28]. Another explanation is that excessive screen time may result in self-isolation, social withdrawal, and the development of internalizing behaviors such as depression and anxiety [15].

We found that only daily screen time ≥ 4h was consistently associated with anxiety, depression, behavior/conduct problems, and ADHD in adjusted logistic regression, which was consistent with another study that used the NSCH2018-2020 dataset and reported a positive association between excessive screen time (≥ 4 hours per day) and behavior/conduct problems and ADHD in both unadjusted and adjusted models [21]. Digital devices can benefit children and adolescents' cognitive outcomes through the quality of interactive content and parent/teacher guidance [29]. Thus, limiting instead of banning screen time, providing guidance on appropriate use of digital devices, and open, non-judgmental discussion to empower children and adolescents to appropriately use digital devices may benefit their mental health [30].

**Mediating effects of physical activity**

Our SEM models consistently found that screen time was directly associated with poorer mental health and ADHD, and indirectly associated with mental health and ADHD through the mediation of physical activity. This finding is in line with other studies [31,32]. Indeed, physical activity can benefit mental health by increasing serotonin and endorphins, regulating and improving mood, distracting children and adolescents from negative thoughts, providing social opportunities, and improving sports performance, which can further boost self-esteem and physical self-perceptions [16]. Prolonged screen use can displace the time for physical activity and other healthy lifestyles and jeopardize child and adolescent mental health [33].

**Mediating effects of sleep**

In this study, simple mediation analyses using the exact natural effects approach and parallel mediation analyses using the SEM method both found that short sleep duration and irregular bedtime mediated the association between screen time and children and adolescent mental health and ADHD. Previous research reported the mediating role of sleep duration and sleep difficulties between screen time and adolescent psychological symptoms, behavior problems, health symptoms, and ADHD symptoms [15,34] while two studies using the pre-pandemic NSCH datasets (NSCH 2018 and NSCH 2019-2020) found sleep duration did not mediate the association between screen time and children and adolescent anxiety and depression [7,8]. One potential explanation for this discrepancy is that our study used the NSCH dataset collected during the pandemic when there was a significant increase in screen time and sleep duration during this period, which may lead to more statistical power to detect the minor mediating role of short sleep duration on the relationship between screen time and mental health. In fact, our study found that the mediating magnitude of short sleep duration was small (ranging between 2.77% to 7.34%), while irregular bedtime explained more variation (ranged 18.2% to 25.7%) of the relationship between screen time and mental health. A previous study also reported that compared to sleep duration, other dimensions of sleep health, such as sleep quality and sleep onset difficulties, had larger mediating effects in the relationship between school pressure and adolescent psychological symptoms [35]. Our study highlighted the significant role of sleep regularity in children's and adolescents' mental health. Future studies should consider wider domains of sleep health, such as regularity, sleep timing, and quality, in improving child and adolescent mental health.

**Multi-group mediation**

Findings from the multi-group mediation analysis underscore the importance of age-specific interventions to address the negative impact of screen time on anxiety, emphasizing the roles of physical activity and regular bedtime routines. Our analysis showed that in the youngest age group (6-10 years), irregular bedtime and short sleep duration significantly mediated the relationship between screen time and depression, with physical activity having a smaller but still significant indirect effect. This finding highlights the crucial role of sleep (both regular bedtime and appropriate duration) in shaping young children's mental health. A potential reason for the relatively less important role of physical activity in the younger children group is that children in this age group are generally physically active. In the 11-13 years and 14-17 years groups, however, physical activity had the strongest effect compared to sleep duration and bedtime regularity. These findings suggest that interventions reducing screen time, promoting physical activity, and ensuring regular bedtimes could be crucial in mitigating mental health problems such as depression among older children and adolescents.

**Limitations**

This study has several limitations. The cross-sectional design prevents us from drawing causal conclusions regarding the relationship between screen time, physical activity, sleep, mental health, and ADHD. Furthermore, screen time, physical activity, and sleep-related variables were reported by parents/caregivers and thus may include self-report bias. In terms of child and adolescent mental health and ADHD, parents reported whether any health provider told the parent/caregiver that the child had the problem. It remains unclear how the diagnosis was made, and the subjective report nature of the outcome variables may not reflect the true mental health status of the child. Despite these limitations, this study has strengths in using national-representative data, including both the individual and family level of covariates in analyses, and using extended Baron and Kenny's approach and SEM approach to highlight the important mediating role of physical activity, sleep duration, and bedtime regularity in the relationship of screen time and mental health and ADHD.

## Public Health Implications

Our findings highlight the need for public health policies that address screen time reduction and support healthy lifestyle behaviors, particularly physical activity and regular bedtime to improve the mental health of children and adolescents. Implementing community-wide programs to encourage physical activity and establish regular sleep patterns could significantly mitigate mental health issues and ADHD.

**Ethical statements**

**Ethical approval:** This analysis of an anonymized public dataset was deemed exempt by the Institutional Review Board (IRB) at the institutions of the first and corresponding authors. The collection of NSCH data was approved by the US National Center for Health Statistics (NCHS) Research Ethics Review Board (ERB) and the Associates Institutional Review Board (IRB), in compliance with HHS regulations (45 CFR 46).

**Informed consent:** Verbal consent was obtained from the parent or caregiver responding to the survey, and this consent was documented in the CATI system. The informed consent script explained the voluntary nature of the survey, assured participants that their responses would remain confidential, and clarified that there would be no penalties for choosing not to answer questions.

**Availability of data and materials:** The dataset(s) supporting the conclusions of this article is(are) available at the United States Census Bureau: https://www.census.gov/programs-surveys/nsch/data/datasets.html.

**Competing interests:** The authors declare that they have no competing interests.

**Funding:** This work was supported by the National Natural Science Foundation of China Youth Scientist Fund (Grant No. 72404063), Guangzhou Municipal Science and Technology Bureau (Grant No. SL2024A04J01923). The funders had no role in the undertaking, data analyses, or reporting of this work.

**Acknowledgements:** None.

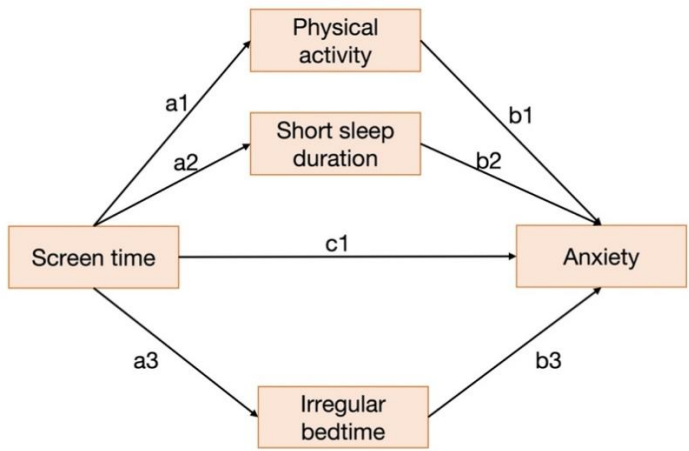

**Figure 1.** Parallel mediation of screen time and anxiety by physical activity, short sleep duration, and irregular bedtime. The three mediators together explained 40.37% of the association between screen time and anxiety. ***: $p < 0.001$.

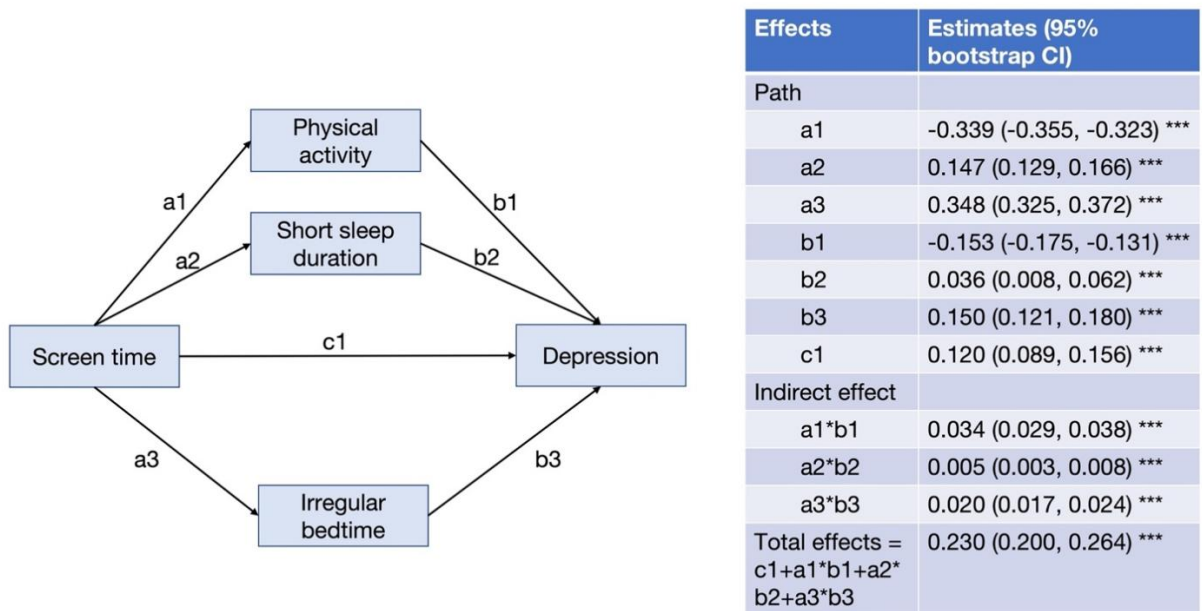

**Figure 2. Parallel mediation of screen time and depression by physical activity, short sleep duration, and irregular bedtime. The three mediators together explained 25.65% of the association between screen time and depression. ***: p < 0.001.**

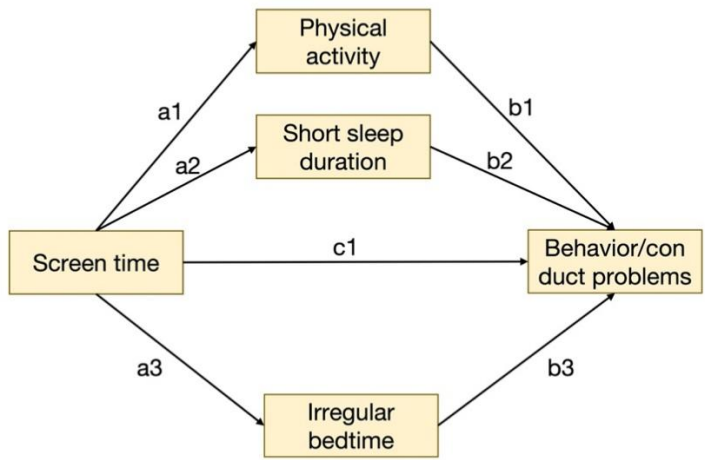

**Figure 3. Parallel mediation of screen time and behavior/conduct problems by physical activity, short sleep duration, and irregular bedtime. The three mediators together explained 54.63% of the association between screen time and behavior/conduct problems. \*\*\*: p < 0.001.**

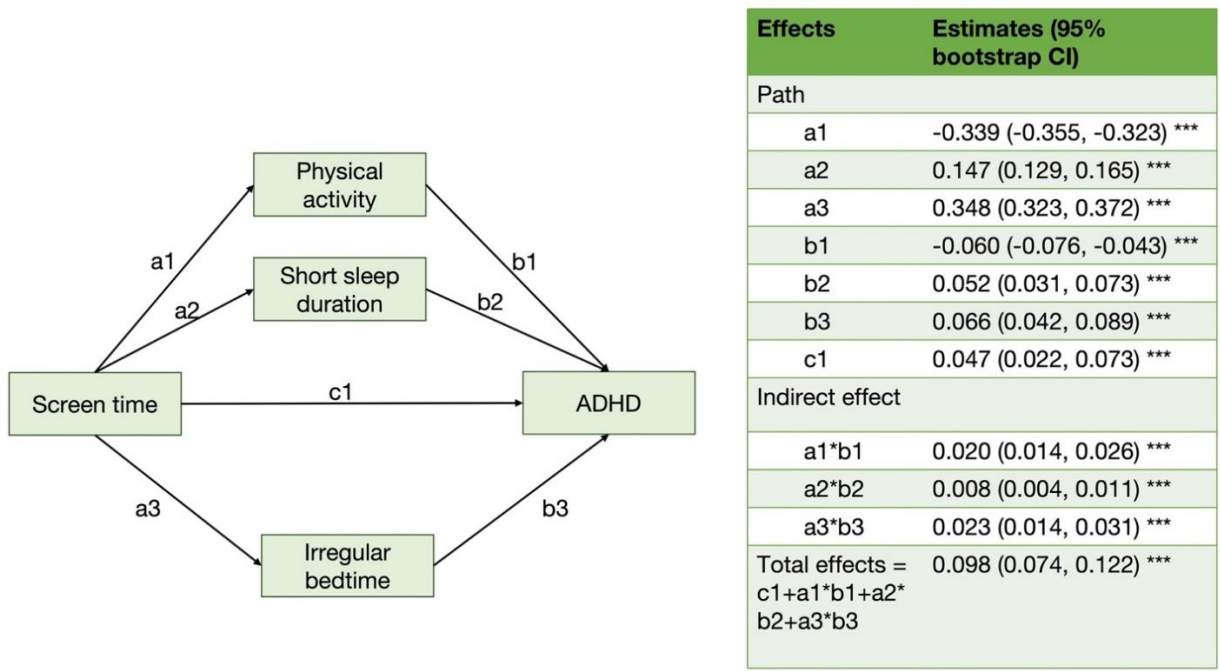

**Figure 4. Parallel mediation of screen time and ADHD by physical activity, short sleep duration, and irregular bedtime. The three mediators together explained 52.04% of the association between screen time and ADHD. *: p < 0.05; **: p < 0.001; ***: p < 0.001.**

**Supplementary Tables**
**Supplementary Table 1. Demographic characteristics of children aged 6-17 years in the NSCH2020-2021 survey (n = 50231)**

| Demographic characteristics | n (%) |
|---|---|
| Child age group | |
|   6-10 yr | 20366 (40.5) |
|   11-13 yr | 8546 (17.0) |
|   14-17 yr | 21319 (42.4) |
| Child gender | |
|   Male | 26049 (51.9) |
|   Female | 24182 (48.1) |
| Child race/ethnicity | |
|   Hispanic | 6521 (13.0) |
|   White, non-Hispanic | 33694 (67.1) |
|   Black, non-Hispanic | 3270 (6.5) |
|   Other/Multi-racial, non-Hispanic | 6746 (13.4) |
| Parent/Caregiver highest education | |
|   Less than high school | 1321 (2.6) |
|   High school | 6567 (13.1) |
|   More than high school | 42343 (84.3) |
| Household income level | |
|   0-99% FPL | 6003 (12.0) |
|   100% - 199% FPL | 8134 (16.2) |
|   200% - 399% FPL | 15389 (30.6) |
|   400% FPL or greater | 20705 (41.2) |
| Family resilience and connection index | |
|   Mean (SD) | 8.90 (2.77) |
|   Median [Min, Max] | 9.00 [0, 12.0] |
| Survey year | |
|   2020 | 25422 (50.6) |
|   2021 | 24809 (49.4) |
| Family Structure | |
|   Two biological/adoptive parents, currently married | 31872 (63.5) |
|   Two biological/adoptive parents, not currently married | 1518 (3.0) |
|   Two parents (at least one not biological/adoptive), currently married | 2938 (5.8) |
|   Two parents (at least one not biological/adoptive), not currently married | 988 (2.0) |
|   Single mother | 8380 (16.7) |
|   Single father | 2635 (5.2) |
|   Grandparent household | 1411 (2.8) |
|   Other relation | 489 (1.0) |
| Digital media use | |
|   Light user (< 2 hours) | 8739 (17.4) |
|   Moderate user (2 - 3 hours) | 25552 (50.9) |
|   Heavy user (>= 4 hours) | 15940 (31.7) |
| Health Insurance Coverage | |
|   Public only (government assistance) | 10604 (21.1) |
|   Private only | 35276 (70.2) |
|   Private and public | 1977 (3.9) |
|   Not insured | 2374 (4.7) |
| Born in US | |
|   Yes | 48153 (95.9) |
|   No | 2078 (4.1) |
| Primary language at home | |
|   English | 46731 (93.0) |
|   Other | 3500 (7.0) |
| Physical activity | |
|   0 days | 5188 (10.3) |
|   1 - 3 days | 19974 (39.8) |
|   4 - 6 days | 14207 (28.3) |
|   Everyday | 10862 (21.6) |
| Insufficient sleep duration | |
|   No | 34719 (69.1) |
|   Yes | 15512 (30.9) |
| Irregular bedtime | |
|   No | 42655 (84.9) |
|   Yes | 7576 (15.1) |
| Current anxiety | |
|   No | 42959 (85.5) |
|   Yes | 7272 (14.5) |
| Current depression | |
|   No | 47040 (93.6) |
|   Yes | 3191 (6.4) |
| Current Behavioral or Conduct problems | |
|   No | 45755 (91.1) |
|   Yes | 4476 (8.9) |
| Current ADHD | |
|   No | 43854 (87.3) |
|   Yes | 6377 (12.7) |

**Supplementary Table 2. Logistic regression of screen time on children's mental health and ADHD**

| Daily Screen time | Anxiety | | Depression | | Behavior/conduct problems | | ADHD | |
|---|---|---|---|---|---|---|---|---|
| | Crude OR (95% CI) | Adjusted OR (95% CI) | Crude OR (95% CI) | Adjusted OR (95% CI) | Crude OR (95% CI) | Adjusted OR (95% CI) | Crude OR (95% CI) | Adjusted OR (95% CI) |
| Light screen time (< 2h) | 1 (Reference) | 1 (Reference) | 1 (Reference) | 1 (Reference) | 1 (Reference) | 1 (Reference) | 1 (Reference) | 1 (Reference) |
| Moderate screen time (2-3h) | 1.17 (1.08, 1.27) *** | 0.99 (0.91, 1.08) | 1.37 (1.26, 1.49) *** | 1.06 (0.97, 1.16) | 1.83 (1.58, 2.12) *** | 1.10 (0.94, 1.28) | 0.96 (0.88, 1.06) | 0.88 (0.80, 0.97) ** |
| Heavy screen time (>= 4h) | 1.86 (1.71, 2.02) *** | 1.21 (1.11, 1.33) *** | 2.63 (2.43, 2.86) *** | 1.45 (1.32, 1.58) *** | 5.11 (4.42, 5.89) *** | 1.65 (1.41, 1.93) *** | 1.61 (1.47, 1.77) *** | 1.17 (1.05, 1.30) *** |

Note: 1. *: $p < 0.05$, **: $p < 0.01$, ***: $p < 0.001$.

2. OR: odds ratio.

3. In all adjusted models, the following variables were adjusted: survey year, age, sex, ethnicity, insurance type, nativity, and primary language spoken at home, parent/caregiver highest education level, household income level, family resilience and connection index (FRCI), short sleep duration, irregular bedtime, and physical activity.

**Supplementary Table 3.** Simple mediation analysis of the association between screen time and mental health and ADHD mediated by physical activity, short sleep duration, and irregular bedtime, respectively, NSCH 2020-21.

| Mediator | Anxiety | | | Depression | | | Behavior/ conduct problems | | | ADHD | | |
|---|---|---|---|---|---|---|---|---|---|---|---|---|
| | OR | SE | 95% boot CI | OR | SE | 95% boot CI | OR | SE | 95% boot CI | OR | SE | 95% boot CI |
| **Physical activity (PA)** | | | | | | | | | | | | |
| Natural direct effect (NDE) | 2.65 | 0.24 | 2.23, 31.9 | 4.4 | 0.63 | 3.39, 5.82 | 1.84 | 0.21 | 1.47, 2.31 | 1.78 | 0.17 | 1.50, 2.16 |
| Natural indirect effect (NIE) | 1.31 | 0.03 | 1.25, 1.36 | 1.5 | 0.04 | 1.42, 1.59 | 1.23 | 0.03 | 1.17, 1.29 | 1.19 | 0.02 | 1.15, 1.24 |
| Total effect | 3.46 | 0.31 | 2.94, 4.12 | 6.6 | 0.93 | 5.11, 8.75 | 2.26 | 0.26 | 1.81, 2.82 | 2.12 | 0.19 | 1.79, 2.54 |
| Proportion mediated by PA | 33.20% | | | 39.30% | | | 33.50% | | | 30.20% | | |
| **Short sleep duration (SSD)** | | | | | | | | | | | | |
| Natural direct effect (NDE) | 3.37 | 0.31 | 2.85, 4.03 | 6.49 | 0.92 | 5.03, 8.59 | 2.1 | 0.24 | 1.68, 2.63 | 2.02 | 0.19 | 1.71, 2.41 |
| Natural indirect effect (NIE) | 1.02 | 0.005 | 1.01, 1.03 | 1.05 | 0.01 | 1.04, 1.07 | 1.03 | 0.01 | 1.02, 1.05 | 1.04 | 0.006 | 1.03, 1.05 |
| Total effect | 3.45 | 0.31 | 2.92, 4.12 | 6.83 | 0.98 | 5.24 | 2.17 | 0.25 | 1.73, 2.71 | 2.1 | 0.19 | 1.78, 2.52 |
| Proportion mediated by SSD | 2.77% | | | 5.58% | | | 5.42% | | | 7.34% | | |
| **Irregular bedtime (IB)** | | | | | | | | | | | | |
| Natural direct effect (NDE) | 3.07 | 0.28 | 2.6, 3.67 | 5.22 | 0.75 | 4.04, 7.01 | 1.94 | 0.22 | 1.55, 2.43 | 1.92 | 0.18 | 1.62, 2.30 |
| Natural indirect effect (NIE) | 1.15 | 0.02 | 1.11, 1.18 | 1.28 | 0.03 | 1.23, 1.35 | 1.15 | 0.02 | 1.11, 1.20 | 1.11 | 0.02 | 1.08, 1.15 |
| Total effect | 3.52 | 0.32 | 2.99, 4.18 | 6.7 | 0.95 | 5.16, 8.89 | 2.24 | 0.25 | 1.79, 2.79 | 2.14 | 0.2 | 1.82, 2.57 |
| Proportion mediated by IB | 18.20% | | | 25.70% | | | 23.60% | | | 18.70% | | |

Note: The proportion mediated by each mediator was calculated using the inverse odds weighting (IOW) approach. All analyzes were adjusted for survey year, child sex, child ethnicity, child nativity, child insurance type, primary language spoken at home, digital media use, household income level, parental highest education level, and FRCI.

**Supplementary Table 4. Chi-square difference tests for multi-group mediation versus constrained model**

|  | CFI | TLI | RMSEA | SRMR | $\Delta\chi^2$ | df | p value for $\Delta\chi^2$ |
|---|---|---|---|---|---|---|---|
| Anxiety |  |  |  |  |  |  |  |
| Full model | 1 | 1 | 0 | 0 |  |  |  |
| Constrained model vs. Full model | 0.899 | 0.870 | 0.029 | 0.011 | 205.146 | 14 | < 0.001 |
| Depression |  |  |  |  |  |  |  |
| Full model | 1 | 1 | 0 | 0 |  |  |  |
| Constrained model vs. Full model | 0.928 | 0.907 | 0.025 | 0.020 | 161.011 | 14 | < 0.001 |
| Behavior/conduct problems |  |  |  |  |  |  |  |
| Full model | 1 | 1 | 0 | 0 |  |  |  |
| Constrained model vs. Full model | 0.924 | 0.902 | 0.024 | 0.013 | 146.211 | 14 | < 0.001 |
| ADHD |  |  |  |  |  |  |  |
| Full model | 1 | 1 | 0 | 0 |  |  |  |
| Constrained model vs. Full model | 0.899 | 0.870 | 0.028 | 0.015 | 191.303 | 14 | < 0.001 |

Note: CFI = Comparative fit index; TLI = Tucker-Lewis index; RMSEA = Root-mean-square error of approximation; SRMR = Standardized root-mean-square residual. df: degrees of freedom. In full models, age group (6-10 years, 11-13 years, and 14-17 years) was entered as the grouping variable. In constrained models, the factor loadings for each age group were set to be equal.

**Supplementary figures**

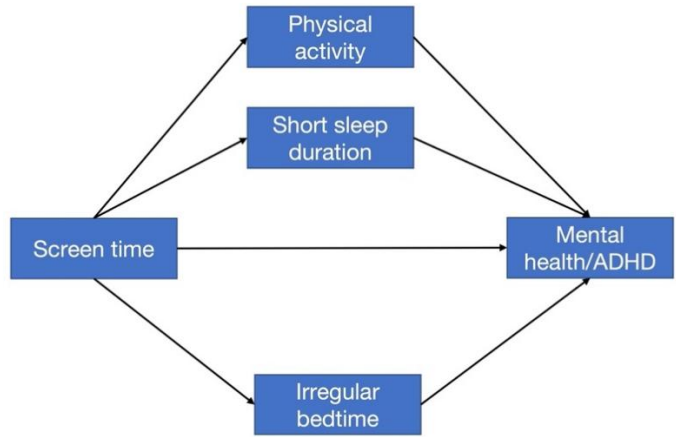

**Supplementary Figure 1. Conceptual model of screen time, physical activity, sleep, and child mental health and ADHD.**

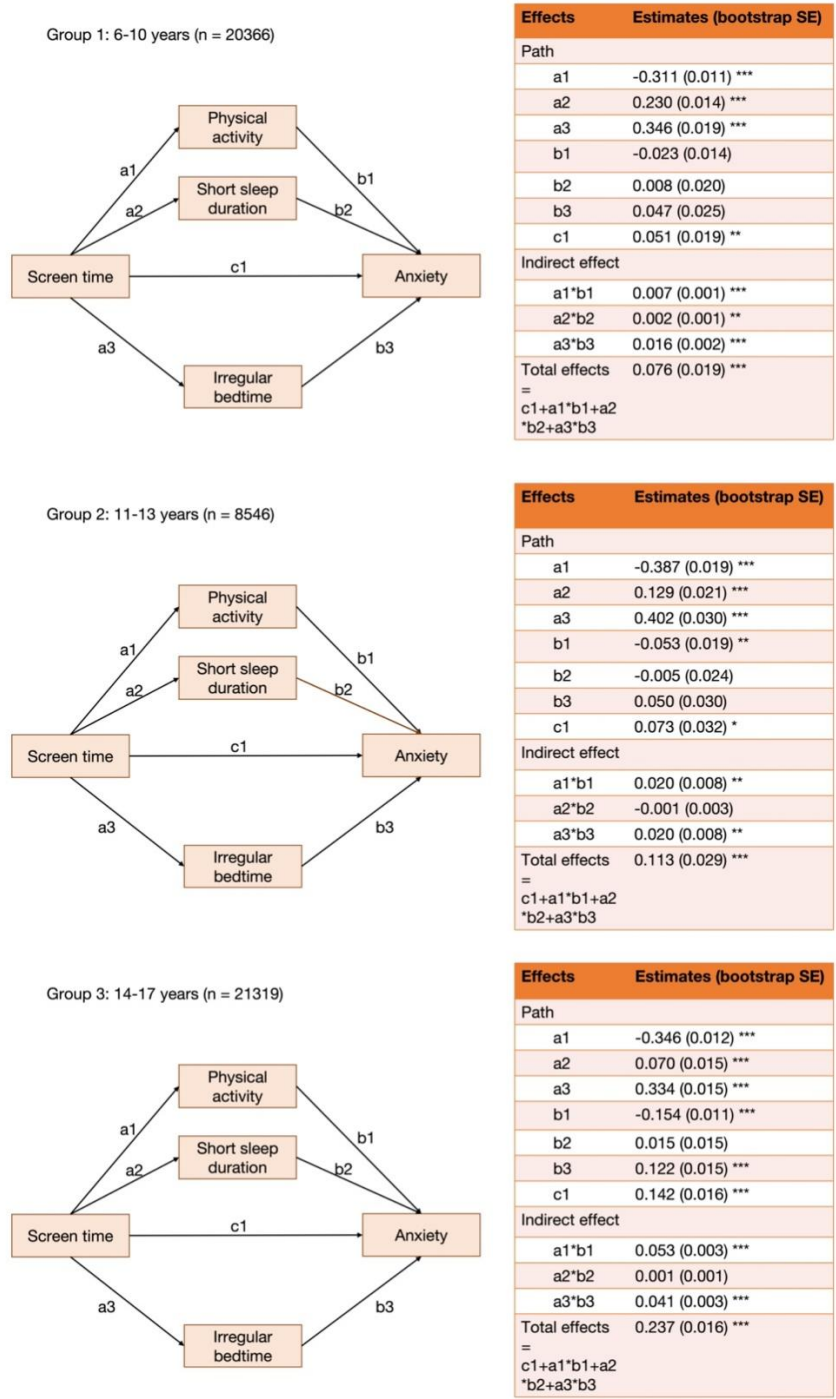

**Supplementary Figure 2. Multi-group mediation of screen time and Anxiety by physical activity, short sleep duration, and irregular bedtime. *: p < 0.05; **: p < 0.001; ***: p < 0.001.**

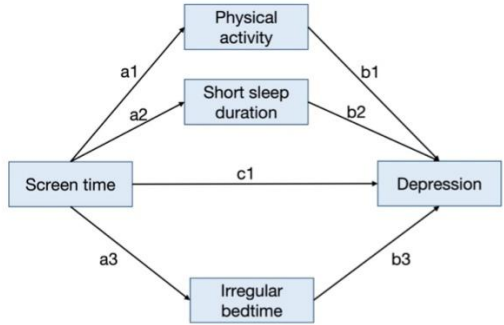

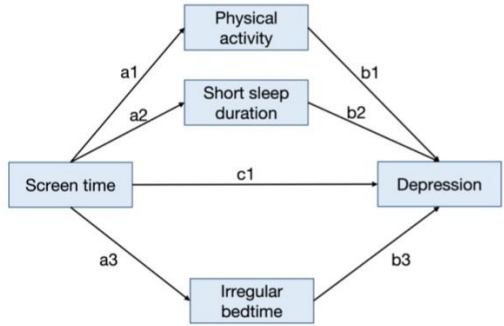

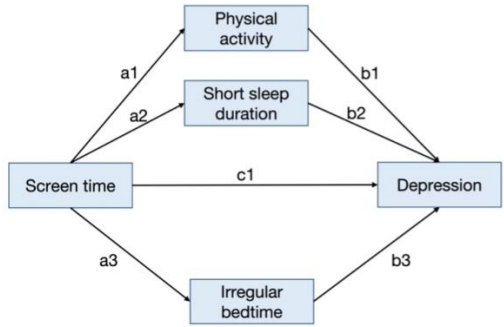

**Supplementary Figure 3. Multi-group mediation of screen time and Depression by physical activity, short sleep duration, and irregular bedtime. \*: p < 0.05; \*\*: p < 0.001; \*\*\*: p < 0.001.**

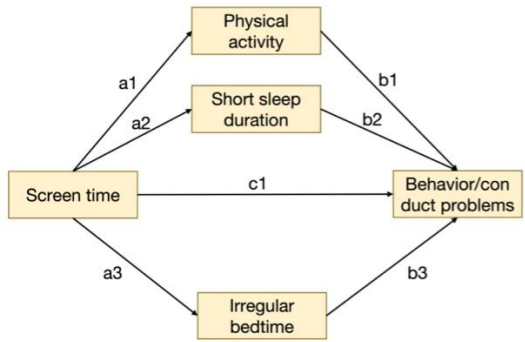

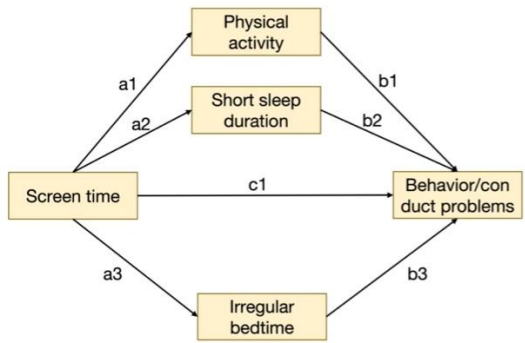

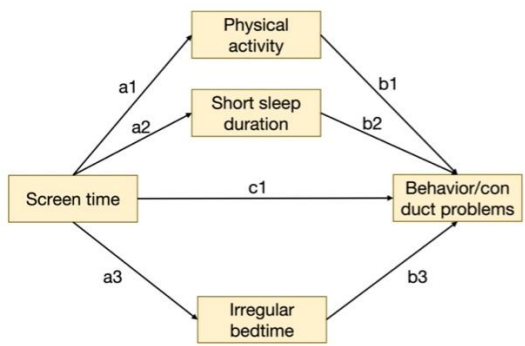

**Supplementary Figure 4. Multi-group mediation of screen time and Depression by physical activity, short sleep duration, and irregular bedtime. *: p < 0.05; **: p < 0.001; ***: p < 0.001.**

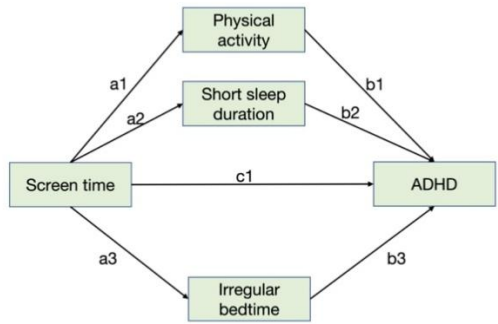

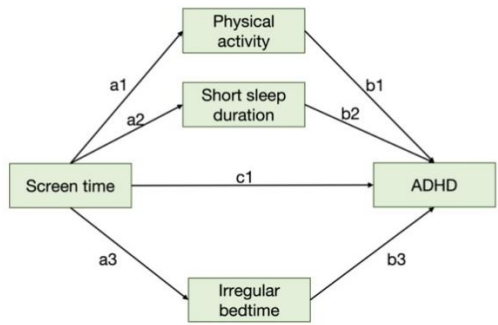

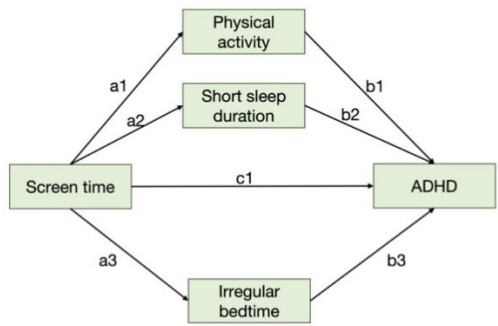

**Supplementary Figure 5. Multi-group mediation of screen time and ADHD by physical activity, short sleep duration, and irregular bedtime. \*: p < 0.05; \*\*: p < 0.001; \*\*\*: p < 0.001.**